\RequirePackage{lineno}
\documentclass[twocolumn,preprintnumbers,amsmath,amssymb,prl,aps]{revtex4}
\usepackage{txfonts}
\usepackage{dsfont}
\usepackage{amssymb}
\usepackage{booktabs}
\usepackage{graphicx}
\usepackage{color}
\usepackage{dcolumn}
\usepackage{amsmath}
\usepackage{epstopdf}
\usepackage{bm}
\usepackage{ulem}
\hfuzz=\maxdimen
\usepackage{physics}

\begin{document}
\title{Direct Measurement of Topological Number by Quench Dynamics}
\author{Pei-Ling Huang$^{1,2}$}
\author{Chao Ma$^{3}$}
\author{Xiang-Long Yu$^{1,4}$}
\email[Corresponding author. E-mail: ]{yuxl@sustech.edu.cn}
\author{Jiansheng Wu$^{1,4,5}$}
\email[Corresponding author. E-mail: ]{wujs@sustech.edu.cn}
\affiliation{1. Shenzhen Institute for Quantum Science and Engineering (SIQSE), Southern University of Science and Technology, Shenzhen, P. R. China}
\affiliation{2. Department of Physics, Southern University of Science and Technology, Shenzhen, P. R. China}
\affiliation{3.  Department of Physics, HuiZhou University, Huizhou, P. R. China}
\affiliation{4. International Quantum Academy (SIQA), Futian District, Shenzhen, P. R. China}
\affiliation{5. Guangdong Provincial Key Laboratory of Quantum Science and Engineering, SIQSE, Southern University of Science and Technology, Shenzhen, P. R. China}

\begin{abstract}

The measurement of topological number is crucial in the research of topological systems.
Recently, the relations between the topological number and the dynamics
are built.  But a direct method to read out the topological number via the dynamics is still lacking.
In this work, we propose a new dynamical protocol to directly measure the topological number of an unknown system.
Different from common quench operations, we change the Hamiltonian of the unknown system to another one with known topological properties.
After the quench, different initial states result in different particle number distributions on the post-quench final Bloch bands. Such distributions depend on the wavefunction overlap between the initial Bloch state and the final Bloch state, which is a complex number depending on the  momentum. We prove a theorem that when  the momentum varies by $2\pi$, the phase of the wavefunction overlap change by $\Delta n\pi$ where $\Delta n$ is the topological number difference between the initial Bloch band and the final Bloch band. Based on this and the known  topological number of the final Bloch band,  we can directly deduce the topological number of the initial state from the particle number distribution and need not track the evolution of the system nor measure the spin texture. Two experimental schemes are also proposed as well. These schemes provide a convenient and robust measurement method and also deepens the understanding of the relation between topology and dynamics.
\end{abstract}

\maketitle

{\it Introduction.}-
The determination of the topological number is important and fundamental in the study of topological phases\cite{ref1,ref2}. Recently, several research groups have reported that quench dynamics can be used to characterize topological properties\cite{ref3,ref4,ref5,ref6,ref7,ref8,ref9,ref10,ref11,ref12,ref13,ref14,ref15,ref16,ref17,ref18,ref19,ref20,ref22,ref23,ref24,ref25}. In these characterizations, the Hamiltonian is changed suddenly and the wavefunction evolves under the post-quench Hamiltonian\cite{ref3,ref4,ref5}. Zhai's group found that the process of the quench dynamics in two-dimensional Chern insulators can be regarded as the mapping of $(k_x,k_y,t)$ on the Bloch sphere\cite{ref17}. If the initial state is topologically trivial, the linking number between two trajectories set by two constraints is equal to the Chern number of the post-quench Hamiltonian. Subsequent researches reveal how to determine the topological invariants in quench dynamics\cite{ref4,ref5,ref6,ref7}.
The study by Liu’s group shows that the topological properties of a post-quench Hamiltonian can be obtained from the winding number defined by the spin texture on the band inversion surface, and it has been successfully observed in ultracold-atom experiments\cite{ref18,ref19}.

To date, most of these dynamical researches focus on the topological characterization of the post-quench Hamiltonian from a known initial state\cite{ref8,ref9,ref10,ref11,ref12,ref13}. Here, we try to find a new dynamical approach for the topological characterization, where we regard the characterized unknown state as the initial state with a given post-quench Hamiltonian.
Our theory shows that the topological number can be obtained from the experimental data by simple procedure instead of certain complicated mathematical processes.
For example, the experiment in Ref. \cite{ref15} needs to track the evolution of the system and some experiments need to integrate the spin texture of the system in the momentum space\cite{ref18,ref19}.
In this paper, a method using quench dynamics to directly measure the topological number of an unknown system is provided.

Our setting of the problem is as follows: assume that a system is prepared in an unknown state. For the sake of simplicity, we consider a two-band system of which  the high energy band is empty and the low energy band is fully occupied (i.e. the unknown state). We measure the topological number of this system through quench dynamics, where the parameters or the external fields are adjusted to obtain a known target Hamiltonian. At this time, the high and low energy bands of the given post-quench Hamiltonian have different particle number distributions and the distributions depend on the unknown initial state. Then particles on the high energy band decay to the low energy band by spontaneous emission of photons\cite{ref35,ref36,ref37}(As illustrated in Fig.1). Through the energy spectrum of the spontaneous emission, we can obtain particle number distribution as well as the topological number of the initial state. In addition, based on our theory, we also propose an experimental scheme in an ultracold-atom system of which the particle number distribution can be obtained directly. We find that in such system, it is easier to obtain the topological number through only measuring atom density as well as the particle number distribution on each energy band.  In the followings, we use a simple two-band model and show the main result numerically, and then give a rigorous proof. Two experimental schemes are provided in the final two sections.

\begin{figure}[tbp]
\centering
\setlength{\abovecaptionskip}{2pt}
\setlength{\belowcaptionskip}{4pt}
\includegraphics[angle=0, width=1 \linewidth]{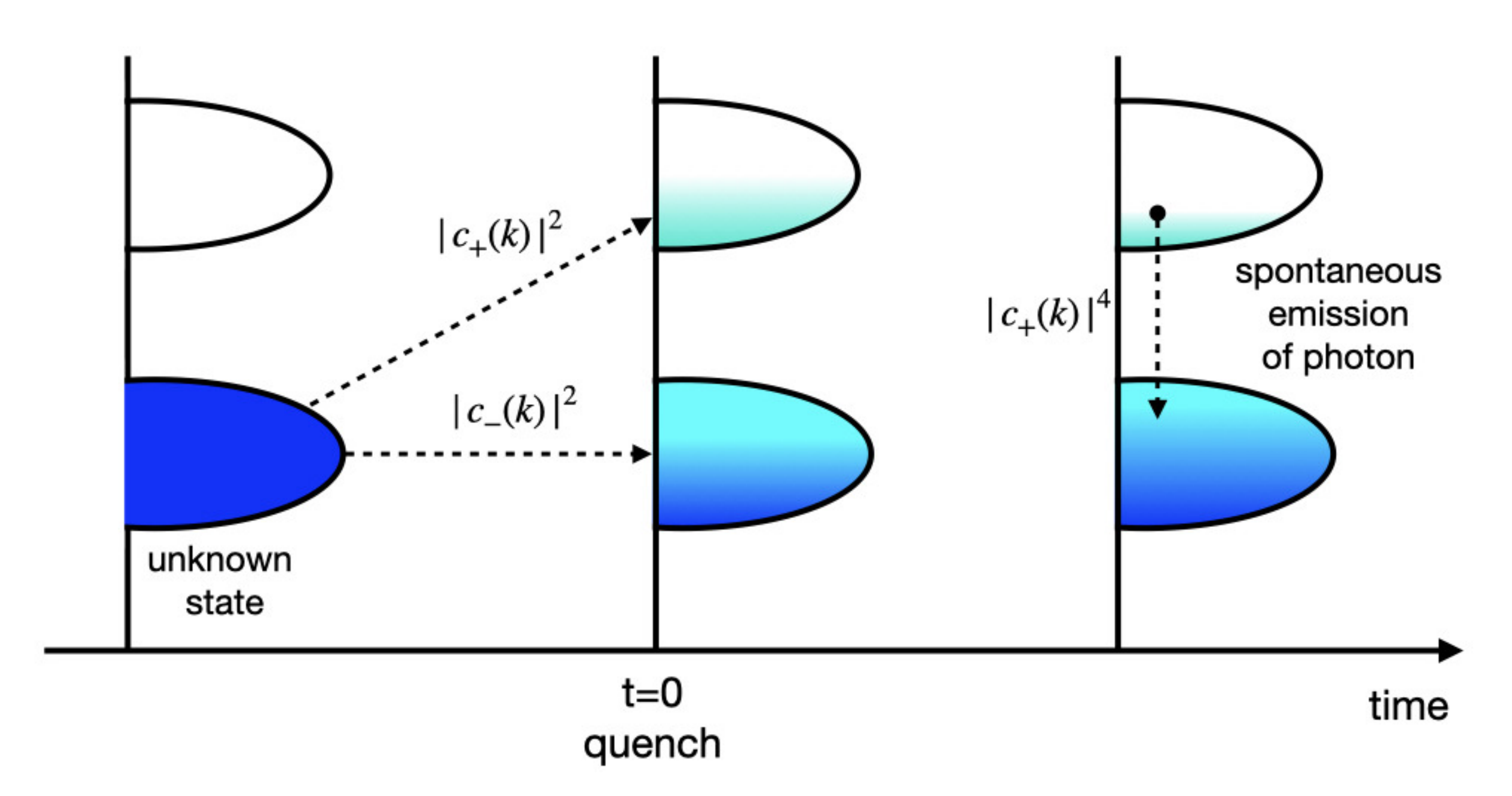}
\caption{The scheme of measurement. An unknown state is projected onto high and low energy bands due to the quench process, then particles on high energy band decay to low energy band by spontaneous emission of photon. The wavefunction overlap $c_+(k)$ is related to the topological number difference of the unknown state and the final state.}\label{Schema}
\end{figure}

{\it Model.}-
Consider a one-dimensional (1D) Qi-Wu-Zhang (QWZ) model whose Bloch Hamiltonian is\cite{ref28b}
\begin{equation}
H=d_z\sigma_z+d_x\sigma_x,\label{model}
\end{equation}
where $d_z(k)=m(t)-2t_s\cos{[n(t)k]},\quad d_x(k)=t_{so}\sin{[n(t)k]}$. Here $m(t)$, $t_s$, $t_{so}$ and $n(t)\in\mathbb{Z}$ are parameters. Setting $t_s, t_{so}>0$ without loss of generality, when $|m(t)|<2t_s$, the system is topologically nontrivial with topological number $\nu=n(t)$, otherwise it is topologically trivial. The eigenvalues of this model are
$
E_{\pm}(k)=\pm\sqrt{d_z^2(k)+d_x^2(k)}.
$
Taking $m(t)>0$, the eigenvectors can be written as
\begin{equation}
\psi_{\pm}(k)
=
\begin{pmatrix}
\cos\theta_{\pm}(k) \\ \sin\theta_{\pm}(k)
\end{pmatrix},
\end{equation}
where $\tan\theta_{\pm}(k)=\pm[\sqrt{1+d_0^2(k)}\mp d_0(k)] d_x(k)/|d_x(k)|$ and $\theta_-(k)=\theta_+(k)+\pi/2$,  $d_0(k)=d_z(k)/|d_x(k)|$.
From this definition, $\theta_-(k)\in [-\pi/2,\pi/2]$, while the global phase of the eigenstates are arbitrary, we choose a gauge such that
$\theta_-(k)$ is continuous in the domain $k\in (-\pi,\pi)$, i. e. the singularity only appear at $k=\pm \pi$. If the system is topologically nontrivial with topological number $n$, $\theta_-(k)$ has a jump of $n\pi$ at $k=\pm \pi$.

We assume that the parameters of Hamiltonian $m(t)=m_1$ and $n(t)=n_1$ at $t<0$,  and after quench $m(t)=m_2$, $n(t)=n_2$ at $t\ge 0$, where
the superscript $1,2$ represent the corresponding physical quantities before and post quench respectively hereafter. Suppose the unknown initial Bloch state is $\ket{\psi^1_-(k)}$. After quench, it is projected onto the final Bloch states, $\ket{\psi^1_-(k)}=c_+\ket{\psi_+^2(k)}+c_-\ket{\psi_-^2(k)}$, we can get
\begin{equation}
\begin{split}
c_{+}(k) &
=-\sin{[\Delta\theta_{-}(k)]}\equiv -\sin{[\theta_{-}^1(k)-\theta^2_-(k)]}, \\
c_{-}(k) &
=\cos{[\Delta\theta_{-}(k)]}.
\end{split}\label{c1c2}
\end{equation}
It can be seen that $c_\pm$ embodies the projection of the initial state on the complete eigenstates of the final Hamiltonian, thus containing the
full information of the initial state.
 $c_{\pm}(k) $ are determined by $\theta_-^{1,2}(k)$.
 From Eq.(\ref{c1c2}), $c_+(k)=0$ when $\Delta\theta_-(k)=0$ or $\pm\pi$, and $c_+(k)=\pm1$ when $\Delta\theta_-(k)=\pm\pi/2$. When $k\rightarrow (2i+1)\pi/n$ $(i\in\mathbb{Z})$, $d_0\rightarrow+\infty$, so $\theta_-\rightarrow\pm\pi/2$.
When $k\rightarrow 2i\pi/n$, there are two cases for the value of $d_0$.
\begin{equation}
\left\{
\begin{aligned}
m(t)>2t_s,\quad d_0\rightarrow+\infty&,\quad\theta_-\rightarrow\pm\pi/2 \\
m(t)<2t_s,\quad d_0\rightarrow-\infty&,\quad\theta_-\rightarrow0
\end{aligned}
\right.
\end{equation}

\begin{figure}[tbp]
\centering
\setlength{\abovecaptionskip}{2pt}
\setlength{\belowcaptionskip}{4pt}
\includegraphics[angle=0, width=1 \linewidth]{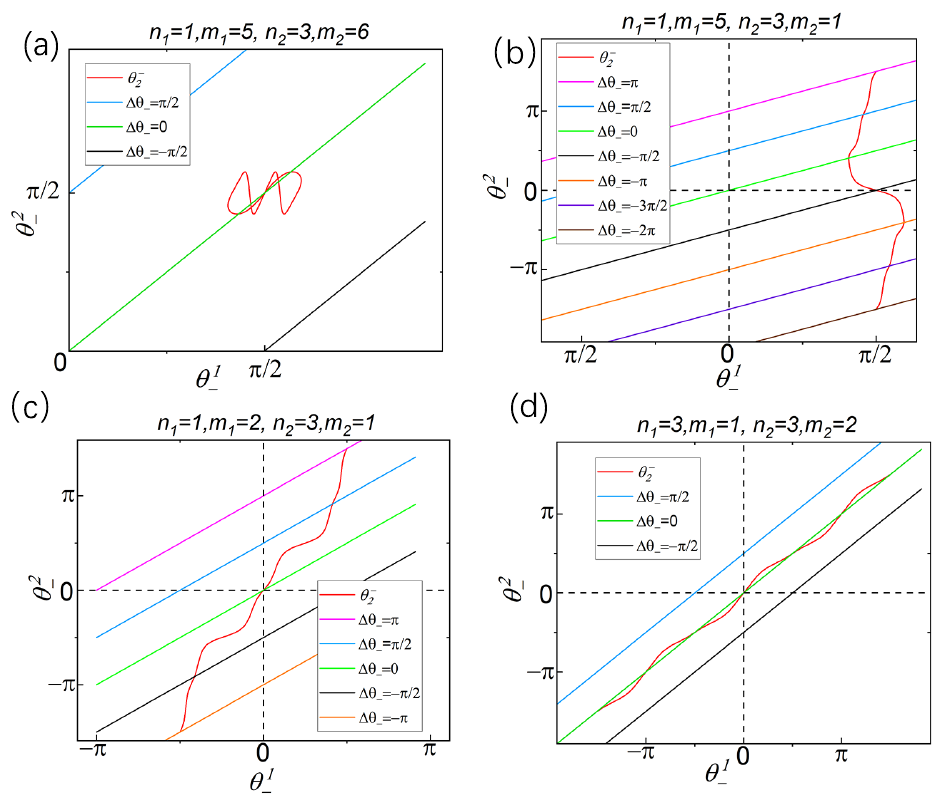}
\caption{$\theta_-^1(k)$ vesus $\theta_-^2(k)$ as $k$ varies from $-\pi$ to $\pi$ for four different cases ($t_s=2$, $t_{so}=1$ for all cases, $n_{1,2}$ and $m_{1,2}$ are indicated on each figures). Different values of $\Delta \theta_-=-\pi,-\pi/2,0,\pi/2,\pi$ are also illustrated.} \label{thetaf}
\end{figure}

\begin{figure}[tbp]
\centering
\setlength{\abovecaptionskip}{2pt}
\setlength{\belowcaptionskip}{4pt}
\includegraphics[angle=0, width=1 \linewidth]{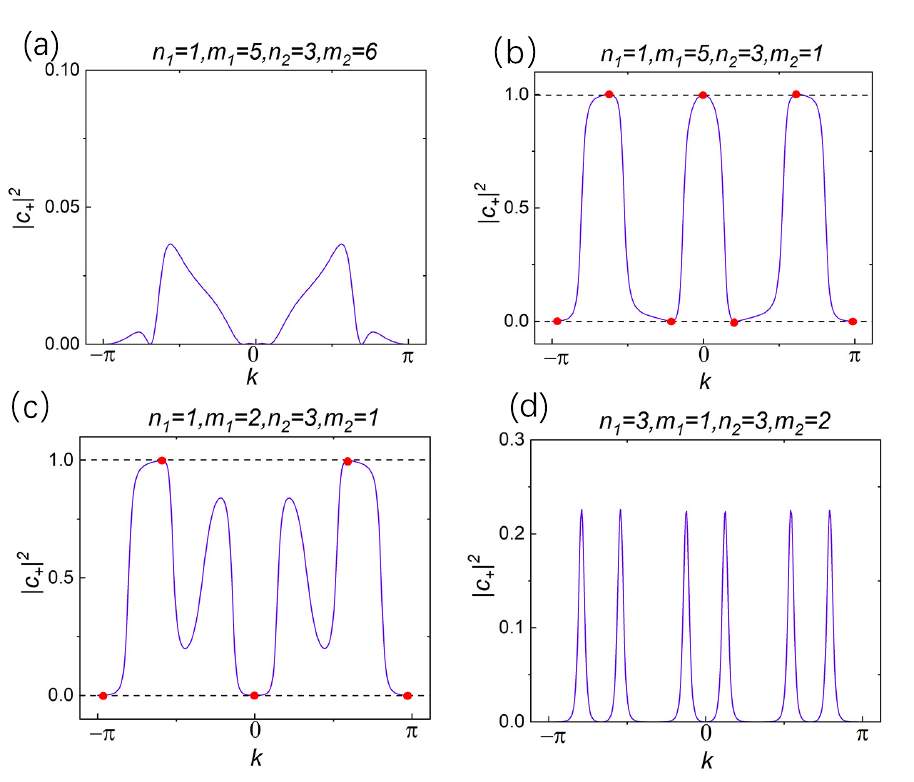}
\caption{ $\abs{c_+(k)}^2$ as $k$ varies from $-\pi$ to $\pi$ for four different cases ($t_s=2$, $t_{so}=1$ for all cases, $n_{1,2}$ and $m_{1,2}$ are indicated on each figures). } \label{cztso1}
\end{figure}

We would discuss $\Delta \theta_-(k)$ as well as $c_+(k)$ in four situations (as illustrated in Fig.\ref{thetaf}).
 (a) $m_{1,2}>2t_s$, i.e. both initial state and final state are topologically trivial. Because $\theta^i_-(k)\ (i=1,2)$ is a continuous function on the interval $(-\pi/n_i,0)$, $(0,\pi/n_i)$, and $d_0>0$, so $\abs{\theta^i_-(k)}>\pi/4$ and $\abs{\Delta \theta_-(k)}<\pi/2$ which means $\abs{c_+(k)}<1$.
(b) $m_1>2t_s$ and $m_2<2t_s$, or $m_1<2t_s$ and $m_2>2t_s$, i.e. one of the initial state and the final state is topologically trivial and the other is topologically nontrivial. If $m_1>2t_s$ and $m_2<2t_s$, $\theta_-^1$ coordinate of the $\theta_-^2(k)$ curve go back to the original phase  as $k$ varies from $-\pi$ to $\pi$ due to its trivial topology and $\theta_-^2(k)$ curve intersect with  $\Delta\theta_-=n\pi\ (n\in {N})$  $n_2+1$ times and intersect with $\Delta\theta=(n-1/2)$  $n_2$ times, thus there exist $n_2$ complete phase cycle (i.e. $n_2\pi$) as $k$ varies from $-\pi$ to $\pi$.
(c) $m_{1,2}<2t_s$ and $n_1\neq n_2$, i.e both initial state and final state are topologically nontrivial with different topological numbers.
Compared with the above case, $\theta_-^2$ coordinate of the upper endpoint of the curve $\theta_-^2(k)$ is shifted to the left by $n_1\pi$, the number of the complete phase cycle is  reduced by $n_1$, thus it is $|n_2-n_1|$.
(d) $m_{1,2}<2t_s$ and $n_1=n_2$, i.e both initial state and final state are topologically nontrivial with same topological numbers.
It is the same case as previous one with $n_1=n_2$, so the number of complete phase cycle is zero. Obviously $\abs{c_+(k)}<1$ in this case.
In summary, for a complete phase cycle of $\Delta\theta_-(k)$,  $c_+^2 (k)$ varies from $0$ to $1$ and back to $0$ which can be defined as a complete peak (CP), then the number of CPs of $c_+^2 (k)$ as $k$ varies from $-\pi$ to $\pi$ is equal to the topological number difference between the initial and final state, as shown in FIG. \ref{cztso1}.
This is the main theoretical results basing on which we can measure the topological number directly.

{\it Proof.}-
Before introducing the experimental schemes, we verify that the above theory is broadly applicable to one-dimensional two-band systems.
Suppose the unknown initial two-band state is $\ket*{\psi_i^{+}}$, we suddenly change the Hamiltonian of the system to  $H_f$ and its eigenstates are  $\ket*{\psi_f^{\gamma}}$ ($\gamma=\pm$). The eigenstates of any two-band model can be written as
\begin{equation}
\ket*{\psi_{l}^+}=\begin{pmatrix}
\cos{\theta_{l}(k)}e^{i\alpha_{l}(k)} \\ \sin{\theta_{l}(k)}e^{i\beta_{l}(k)}
\end{pmatrix},\quad
\ket*{\psi_{l}^-}=\begin{pmatrix}
\sin{\theta_{l}(k)}e^{-i\beta_{l}(k)} \\ -\cos{\theta_{l}(k)}e^{-i\alpha_{l}(k)}
\end{pmatrix}.
\end{equation}
where $l=f,i$. Then, the wavefunction overlap is in forms of
\begin{equation}
\begin{split}
&\abs{\bra*{\psi_f^{+}(k)}\ket*{\psi_i^{-}(k)}}^2 = \sin^2{[\theta_f(k)-\theta_i(k)]}+\sin{[2\theta_f(k)]}
\\ &\times \sin{[2\theta_i(k)]}\sin^2 { \frac{1}{2}{ \left[ (\alpha_i(k)-\beta_i(k))-(\alpha_f(k)-\beta_f(k))\right]  } }.
\end{split}
\end{equation}
When these states are topologically nontrivial,  the Hamiltonian should be in forms of $H(k)=d_i(k)\sigma_i+d_j(k)\sigma_j$ (where $\sigma_{i,j}$ are Pauli matrix, $i,j=x,y,z$) which contain only two of the three Pauli matrices.  Su-Schrieffer-Heeger (SSH) modell, 1D QWZ model and Kitaev chain all belong to this case\cite{ref28a,ref28b,ref28c}.  Thus the eigenstates  have the property that either $\alpha$ and $\beta$ are constant independent of $k$ or $\theta$ is constant independent of $k$ and $\sin^2(2\theta)=1$. It can be proved that (details are provided in the Supplemental Material~\cite{Supplemental})
$\abs{\bra*{\psi_f^{+}(k=k_0+2\pi)}\ket*{\psi_i^{-}(k=k_0+2\pi)}}^2$ can be written in the following form,
\begin{equation}
\begin{split}
\abs{\bra*{\psi_f^{+}(k=k_0+2\pi)}\ket*{\psi_i^{-}(k=k_0+2\pi)}}^2= \\ \sin^2{[\phi_f(k_0)-\phi_i(k_0)+(n_f-n_i)\pi]},
\end{split}
\end{equation}
where $\phi_{f,i}(k)=\theta_{f,i}(k)$ or $[\alpha_{i,f}(k)-\beta_{i,f}(k)]/2$ and  $n_{i}$ and $n_{f}$ are the topological numbers of the initial and final states respectively. We can see from above that when $k$ goes through a period of $2\pi$, the wavefunction overlap $\abs{\bra*{\psi_f^{+}(k)}\ket*{\psi_i^{-}(k)}}^2$ has $\abs{n_i-n_f}$ transitions from $0$ to $1$ and then to $0$ (CP). Obviously, the previous model fits this result. In the followings, we propose two experimental schemes to do the measurement.

{\it Measurement by spontaneous emission.}-
From above we can see counting the number of CPs in $\abs{c_\pm (k)}^2$ can give rise to the topological number difference $\Delta n$ between the initial Bloch state and the final Bloch state. If the topological number of final Bloch state is known,  to obtain the topological number of the unknown initial Bloch state, we just need to measure $\abs{c_\pm (k)}^2$. Right after quench, the particle number distribution is $\abs{c_\pm (k)}^2$, and then the system automatically approaches to equilibrium by spontaneous emission. From the spectrum of the spontaneous emission, $\abs{c_\pm (k)}^2$ as well as the topological properties of the initial state can be obtained.

The radiation intensity for a $+$ state transited to a $-$ state by spontaneous emission is  $I(k)\propto\hbar\omega(k) A_{-+}$, $A_{-+}$ is spontaneous emission coefficient\cite{ref35,ref36,ref37}
\begin{equation}
A_{-+}=\frac{4e^2\omega^{3}}{3\hbar c^3}|\vb*{r}_{-+}|^2,
\end{equation}
where $\omega(k)=(E_+(k)-E_{-}(k))/\hbar$, $\vb*{r}_{-+}=\bra{\psi_-}\vb*{r}\ket{\psi_+}$.
For the final-state Hamiltonian with
$m(t)=m_2$
and
$n(t)=n_2$
in Eq.(1),
\begin{equation}
\vb*{r}_{-+}=\bra{\psi_-}\vb*{r}\ket{\psi_+}=\frac{2nd_xd_zt_{so}\cos{(n_2k)}-4nd_x^2t_s\sin{(n_2k)}}{\omega(k)^2|d_x|},\label{rnn}
\end{equation}
The radiation intensity is $I(k)=\hbar\omega(k) A_{-+}\abs{c_+(k)}^4$
for our model, where $\abs{c_+(k)}^4$ factor is the probability, $+$ state is occupied, and $-$ state is empty.
There are different $k$ corresponding to the same $\omega$, and the conversion of $I(k)$ to $I(\omega)$ requires superimposing different $k$ which corresponding to the same $\omega$. When the energy spectrum of the final state Hamiltonian has only one minimum point, $c_+$ can be obtained from the radiation intensity, thereby deducing the topological number of the initial state Hamiltonian.

\begin{figure}[tbp]
\centering
\setlength{\abovecaptionskip}{2pt}
\setlength{\belowcaptionskip}{4pt}
\includegraphics[angle=0, width=1 \linewidth]{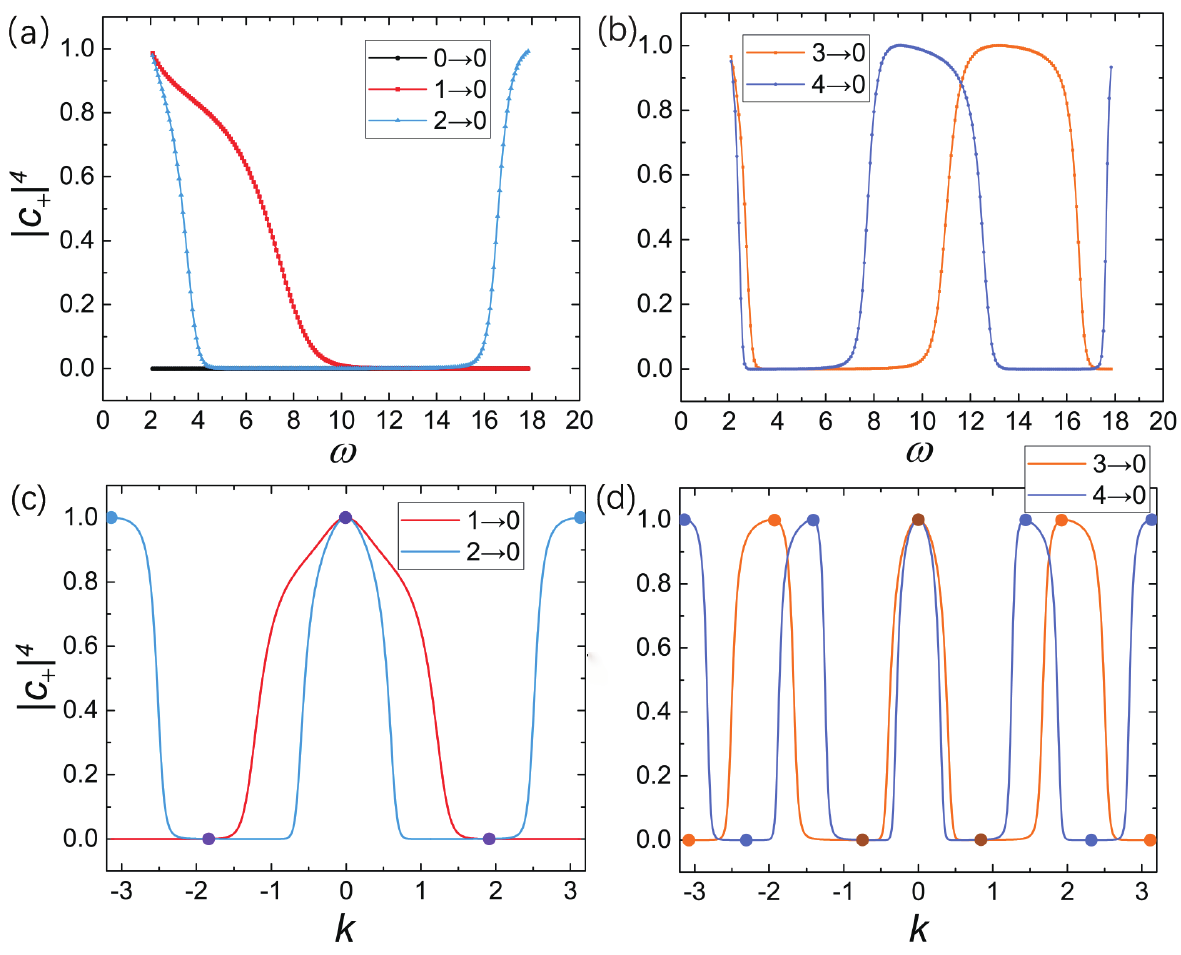}
\caption{$\abs{c_+}^4$ as a function of $k$ and $\omega$. Setting $t_s=2$, $t_{so}=1$, $m_1=1$, and $m_2=5$, ~``$3\rightarrow0$"means that the topological number of initial state is 3 while that of final state is 0. (a)(b)$\abs{c_+(\omega)}^4$ deduced from the radiation intensity, (c)(d)$\abs{c_+(k)}^4$ calculated by analysis.}\label{I_c1}
\end{figure}

\begin{figure}[tbp]
\centering
\setlength{\abovecaptionskip}{2pt}
\setlength{\belowcaptionskip}{4pt}
\includegraphics[angle=0, width=1 \linewidth]{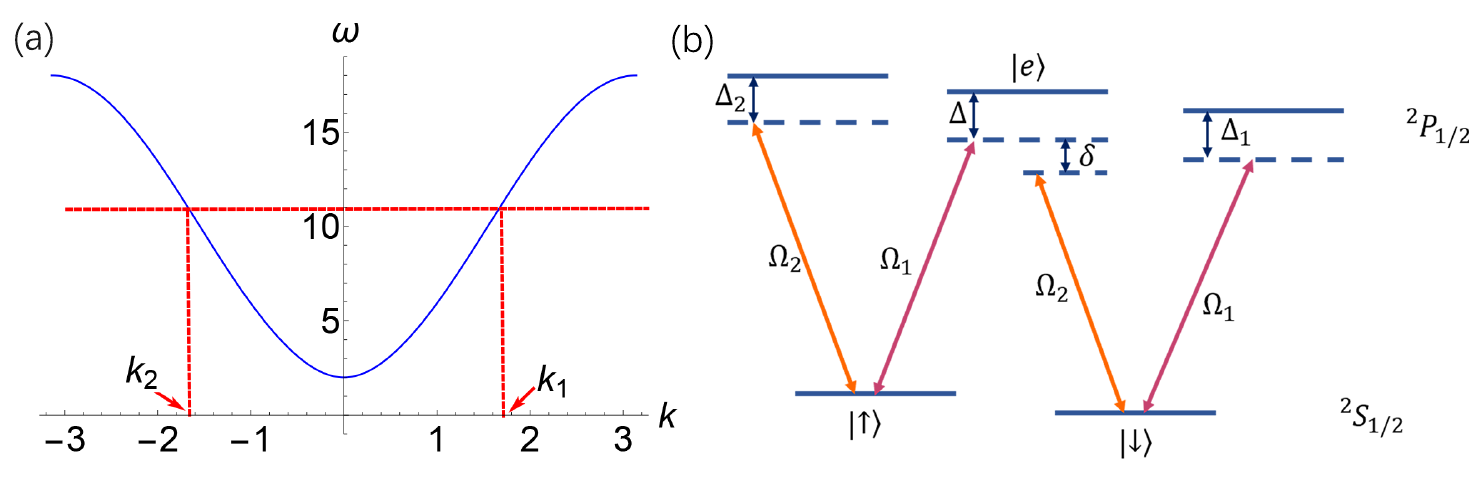}
\caption{(a) Energy spectrum of final Bloch states. $E_-(k_1)=E_-(k_2)$ and $t_s=2$, $t_{so}=1$, $n=1$, $m=5$.
(b) $^{40}\text{K}$ transition induced by the laser field $\Omega_{1,2}$.}\label{enex}
\end{figure}

Taking the final state Hamiltonian $n_2=1$, $m_2=5$ to reduce the complexity, the energy gap $\omega(k)$ is shown in the FIG. \ref{enex}(a), and the same $\omega$ value corresponds to two $k$ values which are symmetric about 0, which is recorded as $k_{1,2}$, and $k_2=-k_1$. From Eq(\ref{c1c2}) and Eq(\ref{rnn}), $c_+(-k)=c_+(k)$, $\vb*{r}(k)=\vb*{r}(-k)$, the radiation intensity $I(\omega)=I(k_1)+I(k_2)=2I(k_1)$.
Then we can obtain
\begin{equation}
\abs{c_+(\omega)}^4\propto I(\omega)\frac{|d_x|^2}{(4d_xd_zt_{so}\cos{k_1}-4d_x^2t_s\sin{k_1})^2}.
\end{equation}
The peak value of $\abs{c_+(\omega)}^4$ reflects half of the peak information of $c_+(k)$. Since $c_+(-k)=c_+(k)$, the topological number of the initial state is equal to 2 times of the number of CPs of $\abs{c_+(\omega)}^4$, as shown in FIG. \ref{I_c1}.

{\it Experimental scheme in cold atom system.}-
The ultracold atoms in optical lattice can simulate the condensed matter Hamiltonian in a clean and well-controlled environment, even when the typical condensed matter system is inaccessible\cite{ref26,ref27,ref28}. It is also suited to investigate topological systems and their properties\cite{ref29,ref30,ref31,ref32,ref33,ref34}.

The above considered model can be simulated by the one-dimensional optical lattice proposed in Ref.\cite{ref18}, which is based on ultracold fermions trapped in an optical lattice, as shown in FIG.\ref{enex}.
Two lasers with Rabi frequencies $\Omega_1(x)=\Omega_0\sin{(k_0x/2)}$ and $\Omega_2(x)=\Omega_0$ are used to induce the two-photon Raman transition: $\ket{\uparrow}, \ket{\downarrow}\rightarrow\ket{e}$. $\Omega_{1,2}$ can also couple respectively to the states $\ket{\uparrow}$ and $\ket{\downarrow}$, but it only leads to additional optical potentials $V\approx-\hbar(\Omega_0^2/\Delta)\sin^2{(k_0x)}$. The one-photon detuning $|\Delta|\gg\Omega_0$ and two-photon detuning $|\delta|\gg\Omega_0$, $|\delta|\ll|\Delta|$. The spin-orbit coupling achieved by this scheme has the following equivalent Hamiltonian
\begin{equation}
H(k)=2t_{so}\sin{(ka)}\sigma_y+(m-2t_s\cos{ka})\sigma_z,
\end{equation}
where $m=\hbar\delta/2$, $t_{so}^{ij}=\int\,dx\phi_{p\uparrow}^{i}(x)(\Omega_0^2/\Delta)\sin{(k_0x)}\phi_{p\downarrow}^{j}(x)=\pm(-1)^jt_{so}^{(0)}$, $t_{so}^{(0)}=(\Omega_0^2/\Delta)\int\,dx\phi_p(x)\sin{(k_0x+\pi/2)\phi_p(x-a)}$, $t_{s}=\int\,dx\phi_{pq}^{j}(x)[p_x^2/2m+V]\phi_{pq}^{j+1}(x)$ ($\phi_{pq}$ means the local $p$ obitals, $q=\uparrow, \downarrow$).
We can directly obtain the topological properties of the initial state by measuring the final atom density. Prepare the system in the initial state $\ket{\downarrow}$, quench the system by suddenly adjusting $\delta$. We only need to measure the atom density $n_{\uparrow\downarrow}({\bf q})$ of each spin component in momentum space by time-of-flight expansion\cite{ref18}, and calculate $c_+^2({\bf q})=n_\uparrow({\bf q})/(n_\uparrow({\bf q})+n_\downarrow({\bf q}))$. Then the topological properties of the initial state can be obtained from the CP number of $c_+^2({\bf q})$.

{\it Conclusion.}-
We have proposed a measurement protocol to directly measure the topological number of one-dimensional two-band system by quench dynamics. By suddenly changing the Hamiltonian of the system to a known one, the initial state of the system projected onto the known target states, which are characterized by the wavefunction overlap of the initial Bloch state and the final Bloch state. On one hand, the wavefunction overlap determines the particle number distribution. On the other hand, we prove that when the momentum varies by $2\pi$, the phase of wavefunction overlap changes by $\Delta n\pi$ where $\Delta n$ is the topological number difference between the initial Bloch band and the final Bloch band. Combining these two aspects and choosing a final Hamiltonian with trivial topology, we can deduce the topological number of the initial Bloch band.
Two experimental schemes are proposed basing on the above result. One is through analyzing the energy spectrum of the spontaneous emission of photon to obtain the magnitude of the wavefunction overlap. The other is to directly measure the particle number distribution in the optical lattice model simulated by ultracold atoms.

From this article and other researches\cite{Sun, Balatsky}, we can see the wavefunction overlap serves as a physical quantity to compare the topological properties of two systems or two states of the same system. For example, it is shown that if the wavefunction overlap of the Bloch states of two insulators is nonzero in the  Brillouin zone, these two insulators can be adiabatically connected\cite{Sun}, which means that they are topologically equivalent. This agrees with our results for the case of $\Delta n=0$. The number of the nodes for the wavefunction overlap is greater than or equal to $\Delta n$\cite{Balatsky}. This also agrees with our results that the number of nodes can be greater than or equal to $\Delta n$ but the number of CPs is equal to $\Delta n$. At the same time, wavefunction overlap is related to dynamics of the system, such as transition amplitude, so it provides many possibilities to obtain the topological properties from the dynamics of the system.
Our proposed protocol is the realization of one of these possibilities.

This protocol does not need to track the evolution of the system, nor does it need to integrate the spin texture of the system in the momentum space.
And the most important feature is that it is not sensitive to the parameter since the number of CPs depends only on the topological number difference between the initial Bloch state and the final Bloch state. So if only the changes of parameters don't change the topological properties,  the results are still robust. It is more efficient and robust and provides another avenue for the future study of topological systems. Our method works for one-dimensional two-band system with topological winding number $Z$, how to extend to higher dimensional systems with three or more bands and to topological system with topological number $Z_2$ are left for future investigation.
\begin{acknowledgments}
This work is supported by National Natural Science Foundation of China (Grants No. U1801661 and No. 11604142), Guangdong Innovative and Entrepreneurial Research Team Program (Grant No. 2016ZT06D348) and Natural Science Foundation of Guangdong Province (Grant No.2017B030308003), and Science, Technology, and Innovation Commission of Shenzhen Municipality (Grants No. JCYJ20190809120203655, No. KYTDPT20181011104202253, No. ZDSYS20170303165926217 and No. JCYJ20170412152620376,). C. Ma is supported by HuiZhou University Dr Fund.
\end{acknowledgments}

\end{document}


\maketitle

\section{Proof}

We consider a one-dimensional (1D) two-band system with an unknown Hamiltonian and the initial state is in the low energy eigenstate 
$\ket*{\psi_i^{-}}$. We suddenly change the Hamiltonian of the system to $H_f$ and its eigenstates are  $\ket*{\psi_f^{\gamma}}$ ($\gamma=\pm$).  Assume that the topological numbers of the initial and final states are $n_{i}$ and $n_{f}$ respectively. We have the eigenstates as
\begin{equation}
\ket*{\psi_{l}^+}=\begin{pmatrix}
\cos{\theta_{l}(k)}e^{i\alpha_{l}(k)} \\ \sin{\theta_{l}(k)}e^{i\beta_{l}(k)}
\end{pmatrix},\quad
\ket*{\psi_{l}^-}=\begin{pmatrix}
\sin{\theta_{l}(k)}e^{-i\beta_{l}(k)} \\ -\cos{\theta_{l}(k)}e^{-i\alpha_{l}(k)}
\end{pmatrix}.
\end{equation}
where $l=f,i$, then the wavefunction overlap is in forms of
\begin{equation}
\abs{\bra*{\psi_f^{+}(k)}\ket*{\psi_i^{-}(k)}}^2=\sin^2{[\theta_f(k)-\theta_i(k)]}+\sin{[2\theta_f(k)]}\sin{[2\theta_i(k)]}\sin^2{\frac{1}{2}[(\alpha_i(k)-\beta_i(k))-(\alpha_f(k)-\beta_f(k))]}.
\end{equation}
When the 1D two-band system is topologically nontrivial,  the Hamiltonian should be in forms of $H(k)=d_i(k)\sigma_i+d_j(k)\sigma_j$ (where $\sigma_{i,j}$ are Pauli matrix, $i,j=x,y,z$) which contain only two of the three Pauli matrices. 1D Qi-Wu-Zhang (QWZ) model, Su-Schrieffer-Heeger (SSH) model and Kitaev chain all belong to this case\cite{ref28a,ref28b,ref28c}.  Thus the eigenstates  have the property that either $\alpha$ and $\beta$ are constant independent of $k$ or $\theta$ is constant independent of $k$ and $\sin^2(2\theta)=1$.
We discuss these two situations respectively:\\

(a) When $\alpha$ and $\beta$ are constant independent of $k$ (either $d_x=0$ or $d_y=0$), 

The detailed Hamiltonian for initial and final states are in forms of  $H(k)=d_\mu(k)\sigma_\mu +d_z(k)\sigma_z$ where $\mu=x$ or $y$, and its eigenvectors are as follows
\begin{equation}
\ket*{\psi_{\pm}(k)}=\begin{pmatrix}
\cos{\theta_{\pm}(k)} \\ \sin{\theta_{\pm}(k)} e^{-i\Theta_\mu}
\end{pmatrix},
\end{equation}
where $\theta_{\pm}=\arctan{(E_{\pm}(k)-d_z(k))/d_\mu (k)}$, $E_{\pm}=\pm\sqrt{d_\mu^2+d_z^2}$ and $\Theta_x=0$ for $\mu=x$ and $\Theta_y=\pi/2$ for $\mu=y$.  Assume that the topological numbers of the system is $n$, which can be expressed as `skew polarization'\cite{skew}
\begin{equation}
\frac{1}{\pi i} \int_{-\pi}^{\pi} \bra*{S_\mu\psi_\pm (k)}\partial_k\ket*{\psi_\pm (k)} \dd k=\frac{1}{\pi i} \int_{-\pi}^{\pi}\partial_k \theta(k) \dd k=n,
\end{equation} 
where $S_x=\sigma_y$ for $\mu=x$ case and $S_y=\sigma_x$ for $\mu=y$ case. The above equation means that when $k$ goes through a period of $2\pi$, $\theta(k)$ vary by $n\pi$.\\

The wavefunction overlap between initial state and final state is in forms of 
$$\abs{\bra*{\psi_f^{+}(k)}\ket*{\psi_i^{-}(k)}}^2=\sin^2{[\theta_f(k)-\theta_i(k)]}.$$ 
When $k$ goes through a period of $2\pi$, $\theta_i(k)$ and $\theta_f(k)$ vary by $n_i\pi$ and $n_f\pi$ respectively, 
\begin{equation}
\abs{\bra*{\psi_f^{+}(k)}\ket*{\psi_i^{-}(k)}}^2_{k=k_0+2\pi}=\sin^2{\left[(\theta_f(k)-\theta_i(k))+(n_f-n_i)\pi \right]}.
\end{equation}
Thus $\abs{\bra*{\psi_f^{+}(k)}\ket*{\psi_i^{-}}(k)}^2$ has $\abs{n_f-n_i}$ transitions from $0$ to $1$ and then to $0$. \\

(b) When $\theta$ is independent of $k$ and $\sin^2(2\theta)=1$ ($d_z=0$) 

The detailed Hamiltonian for initial and final states are in forms of $H(k)=d_x(k)\sigma_x+d_y(k)\sigma_y$, its eigenvectors as follows
\begin{equation}
\ket*{\psi_{+}(k)}=\frac{1}{\sqrt{2}}\begin{pmatrix}
e^{i\alpha(k)} \\ 1
\end{pmatrix},\quad
\ket*{\psi_{-}(k)}=\frac{1}{\sqrt{2}}\begin{pmatrix}
1 \\ -e^{-i\alpha(k)}
\end{pmatrix},
\end{equation}
where $\alpha(k)=\arctan{d_y(k)/d_x(k)}$.  Assume that the topological numbers of the system is $n$, we have
\begin{equation}
\frac{1}{\pi i} \int_{-\pi}^{\pi} \bra*{\sigma_z\psi_\pm(k)}\partial_k\ket*{\psi_\pm (k)} \dd k=\frac{1}{\pi i} \int_{-\pi}^{\pi} \frac{\partial_k \alpha(k)}{2} \dd k=n,
\end{equation} 
when $k$ goes through a period of $2\pi$, $\alpha(k)$ or $\alpha(k)-\beta(k)$ if the wavefunction is in forms of Eq.(1) vary by $2n\pi$.\\

The wavefunction overlap between initial state and final state is in forms of 
$$\abs{\bra*{\psi_f^{+}(k)}\ket*{\psi_i^{-}(k)}}^2=\sin^2{\frac{1}{2} [(\alpha_i(k)-\beta_i(k))-(\alpha_f(k)-\beta_f(k))] }.$$
When $k$ goes through a period of $2\pi$, $\alpha_i(k)-\beta_i(k)$ and $\alpha_f(k)-\beta_f(k)$ vary by $2n_i\pi$ and $2n_f\pi$ respectively, and
\begin{equation}
\abs{\bra*{\psi_f^{+}(k)}\ket*{\psi_i^{-}(k)}}^2_{k=k_0+2\pi}=\sin^2\{\frac{1}{2}\left[(\alpha_i(k)-\beta_i(k))-(\alpha_f(k)-\beta_f(k))\right]+(n_i-n_f)\pi \}.
\end{equation}
Thus $\abs{\bra*{\psi_f^{+}(k)}\ket*{\psi_i^{-}(k)}}^2$ has $\abs{n_f-n_i}$ transitions from $0$ to $1$ and then to $0$ .

In summary, when $k$ varies by $2\pi$, the wavefunction overlap  can be written in the following form,
\begin{equation}
\abs{\bra*{\psi_f^{+}(k=k_0+2\pi)}\ket*{\psi_i^{-}(k=k_0+2\pi)}}^2=\sin^2{[\phi_f(k_0)-\phi_i(k_0)+(n_f-n_i) \pi]},
\end{equation}
where $\phi_{f,i}(k)=\theta_{f,i}(k)$ or $[\alpha_{f,i}(k)-\beta_{f,i}(k)]/2$. So when $k$ goes through a period of $2\pi$, $\abs{\bra*{\psi_f^{+}(k)}\ket*{\psi_i^{-}(k)}}^2$ has $\abs{n_f-n_i}$ transitions from $0$ to $1$ and then to $0$.

\section{Sensitivity to Parameters}

\begin{figure}[htbp]
\centering
\setlength{\abovecaptionskip}{2pt}
\setlength{\belowcaptionskip}{4pt}
\includegraphics[angle=0, width=1 \linewidth]{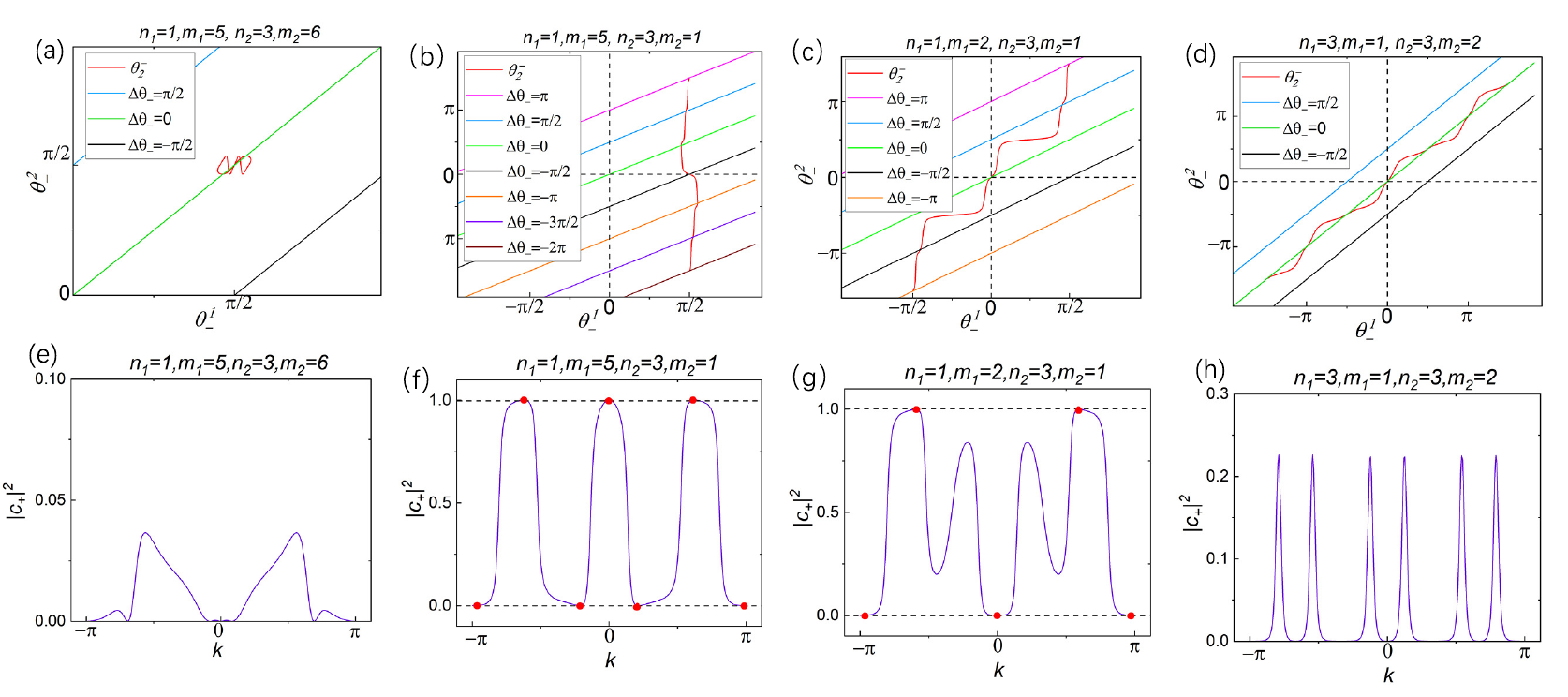}
\caption{Figures of $\theta_-^1(k)$ vesus $\theta_-^2(k)$ and figures of $\abs{c_+}^4 (k)$ for four different cases when parameters are set as $t_s=2$, $t_{so}=1$. (a)(b)(c)(d)the figures of $\theta_-^1(k)$ vesus $\theta_-^2(k)$ as $k$ varies fro $-\pi$ to $\pi$ for four different cases. Different values of $\Delta\theta_-=-\pi,-\pi/2,0,\pi/2,\pi$ are also illustrated; (e)(f)(g)(h)the figures of $|c_+(k)|^2$ as $k$ varies from $-\pi$ to $\pi$ for four different cases. $|c_+(k)|^2=1, 0$ are indicated with red filled circles hereafter.}\label{tso1}
\end{figure}

\begin{figure}[htbp]
\centering
\setlength{\abovecaptionskip}{2pt}
\setlength{\belowcaptionskip}{4pt}
\includegraphics[angle=0, width=1 \linewidth]{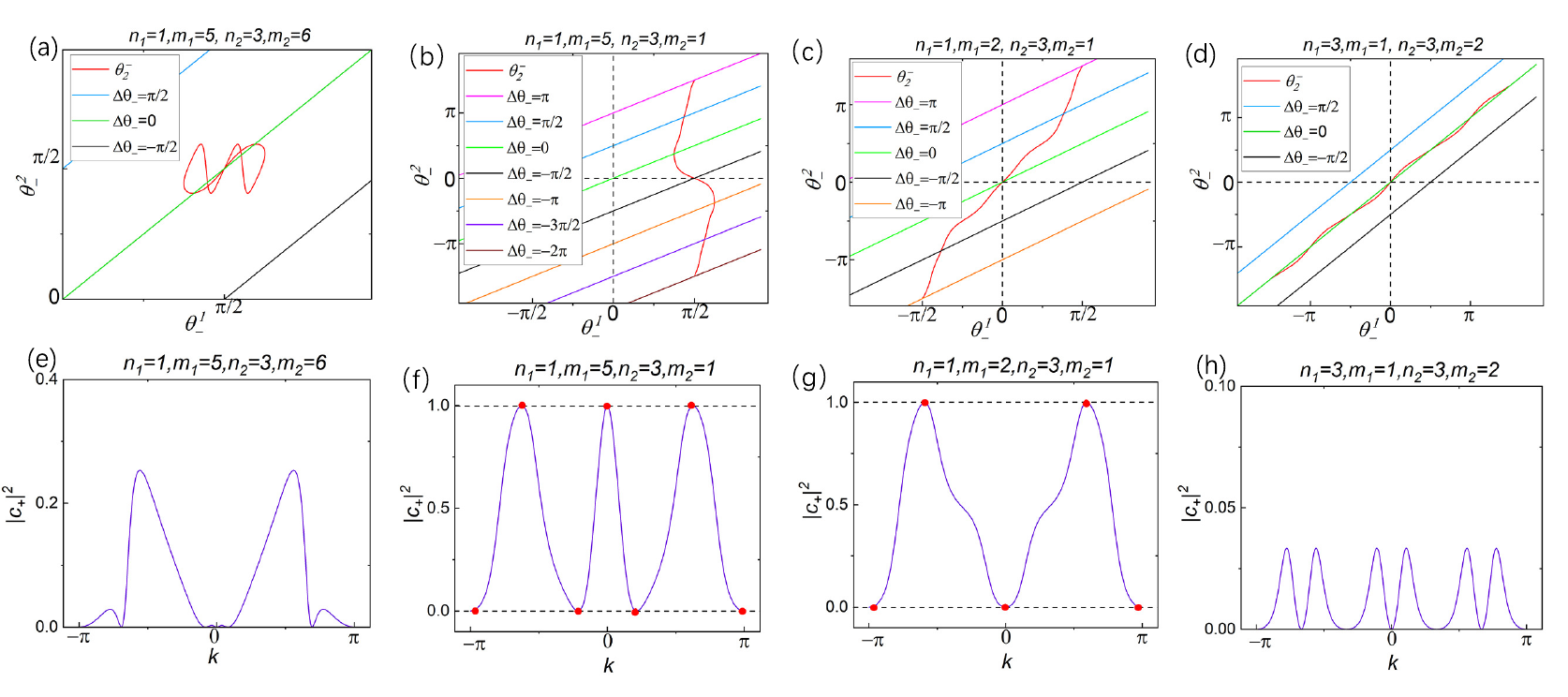}
\caption{Figures of $\theta_-^1(k)$ vesus $\theta_-^2(k)$ and figures of $\abs{c_+}^4 (k)$ for four different cases when parameters are set as $t_s=2$, $t_{so}=3$. (a)(b)(c)(d)the figures of $\theta_-^1(k)$ vesus $\theta_-^2(k)$ as $k$ varies fro $-\pi$ to $\pi$ for four different cases. Different values of $\Delta\theta_-=-\pi,-\pi/2,0,\pi/2,\pi$ are also illustrated; (e)(f)(g)(h)the figures of $|c_+(k)|^2$ as  $k$ varies from $-\pi$ to $\pi$ for four different cases.}\label{tso3}
\end{figure}

\begin{figure}[htbp]
\centering
\setlength{\abovecaptionskip}{2pt}
\setlength{\belowcaptionskip}{4pt}
\includegraphics[angle=0, width=1 \linewidth]{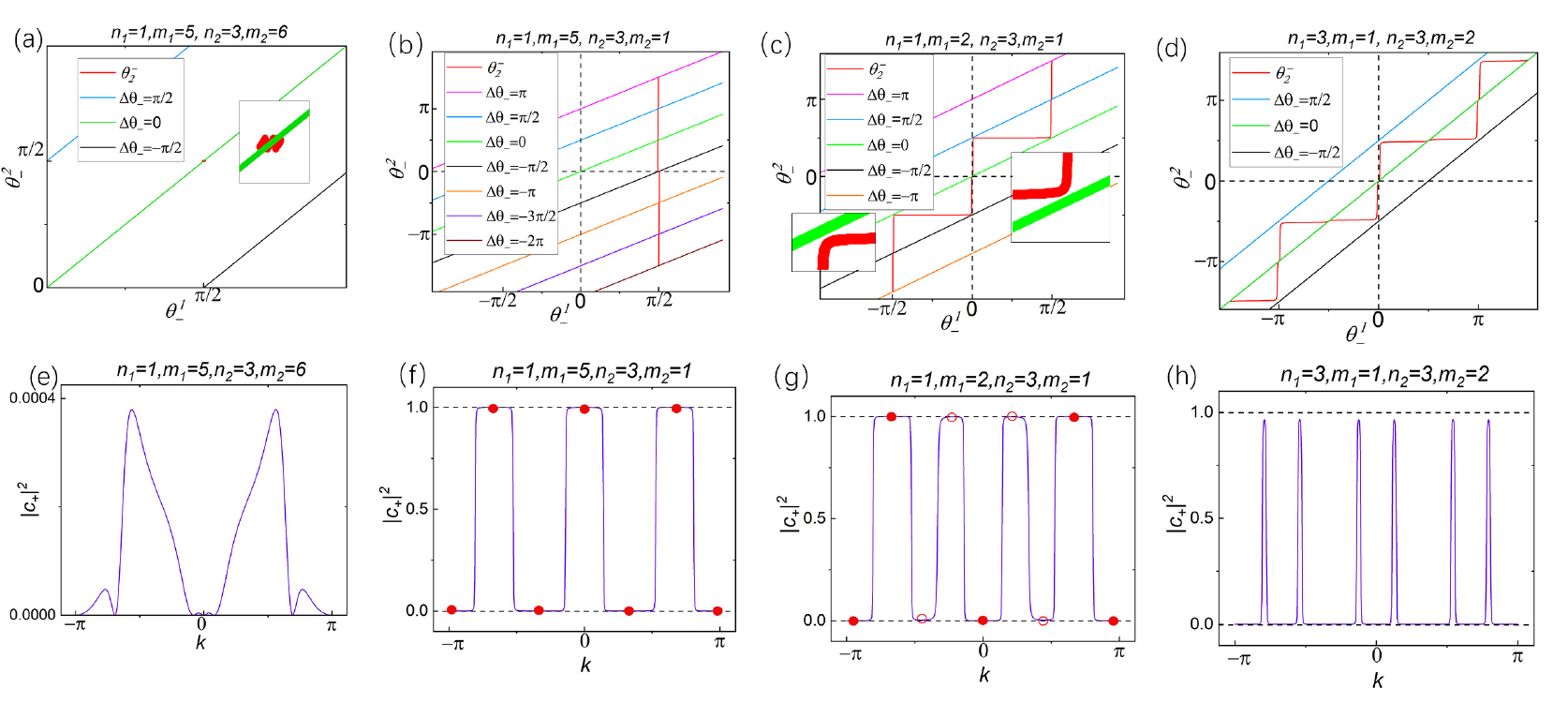}
\caption{Figures of $\theta_-^1(k)$ vesus $\theta_-^2(k)$ and figures of $\abs{c_+}^4 (k)$ for four different cases when parameters are set as $t_s=2$, $t_{so}=0.1$. (a)(b)(c)(d) show the figures of $\theta_-^1(k)$ vesus $\theta_-^2(k)$ as $k$ varies fro $-\pi$ to $\pi$ for four different cases. Different values of $\Delta\theta_-=-\pi,-\pi/2,0,\pi/2,\pi$ are also illustrated; (e)(f)(g)(h) show the figures of $|c_+(k)|^2$ as  $k$ varies from $-\pi$ to $\pi$ for four different cases. $|c_+(k)|^2$ close to but not equal to $1$ are indicated with  red empty circles (false CPs). }\label{tso01}
\end{figure}

We show more examples of different values of parameters for the 1D Qi-Wu-Zhang model mentioned in the main article\cite{ref28b}. When $t_{so}$ is within a certain range, it is easy to count the number of CPs as $k$ varies by $2\pi$. And the number of CPs is not sensitive to the parameters when $t_s$ is not too large or too small compared with $t_s$ (as illustrated in Fig.\ref{tso1}, \ref{tso3}). 
 While when $t_{so}$ is much smaller than $t_s$ (as illustrated in Fig.\ref{tso01}), there are some redundant points $\Delta\theta_-$ close to $\pi/2$, $-\pi/2$ or $0$, causing $|c_+|^2$ to be close to $1$ or $0$. 
 This induces some false CPs, which requires improving the accuracy of the measurement. 
 In this case, we can also chose $n_f=0$ to avoid the case of false CPs.

\section{Another Example: Su-Schrieffer-Heeger Model}
\begin{figure}[htbp]
\centering
\setlength{\abovecaptionskip}{2pt}
\setlength{\belowcaptionskip}{4pt}
\includegraphics[angle=0, width=1 \linewidth]{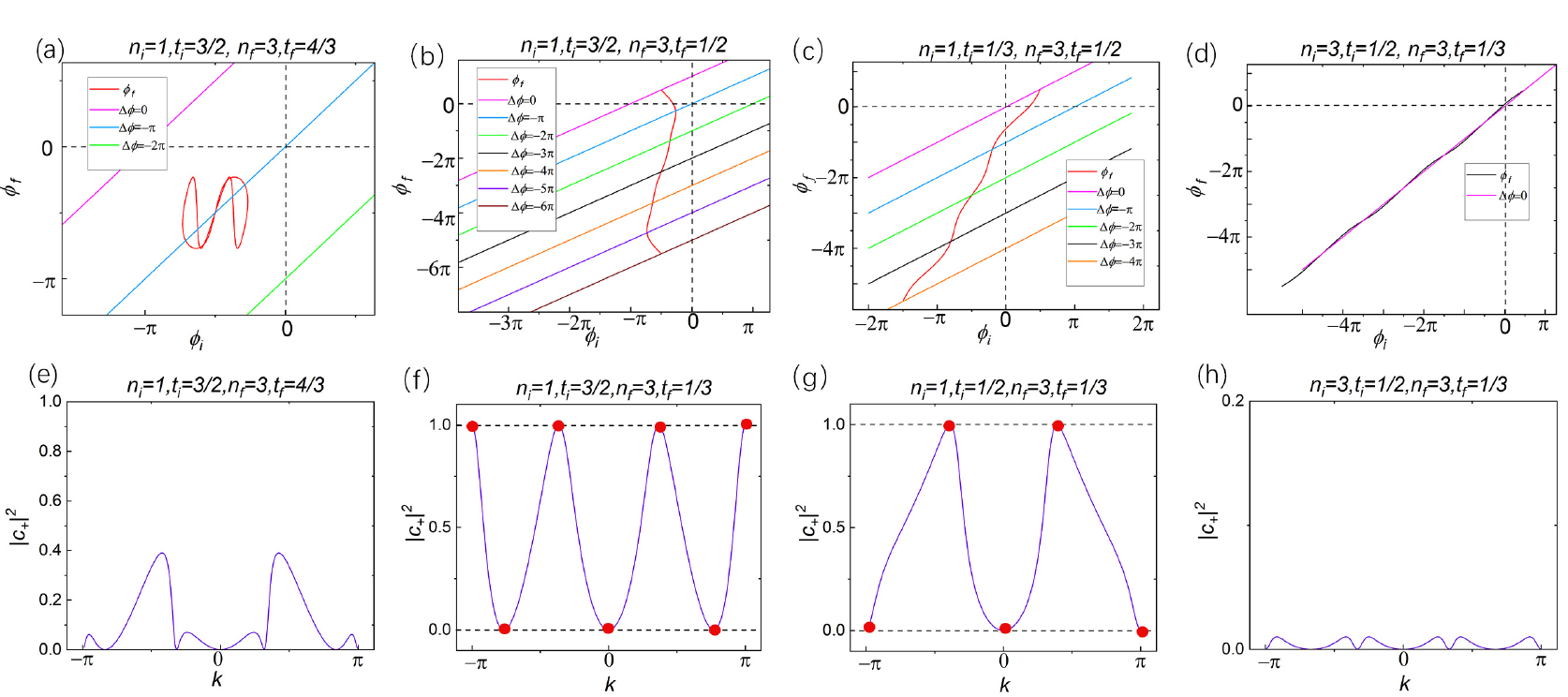}
\caption{Figures of $\theta_-^1(k)$ vesus $\theta_-^2(k)$ and figures of $\abs{c_+}^4 (k)$ for four different cases of SSH model. (a)(b)(c)(d)the figures of $\phi_f(k)$ vesus $\phi_i(k)$ as $k$ varies from $-\pi$ to $\pi$ for four different cases. Different values of $\Delta\phi=0,\pi,2\pi,3\pi,4\pi$ are also illustrated; (e)(f)(g)(h)the figures of $|c_+(k)|^2$ as  $k$ varies from $-\pi$ to $\pi$ for four different cases.}\label{ssh}
\end{figure}

We consider another model, Su-Schrieffer-Heeger (SSH) model\cite{ref28a}. Its Hamiltonian is as follows:
\begin{equation}
H=d_x(k)\sigma_x+d_y(k)\sigma_y,
\end{equation}
where $d_x(k)=t_1(t)+t_2\cos[n(t)k]$, $d_y(k)=t_2\sin[n(t)k]$. When $t_1(t)<t_2$, the system is topological nontrivial and its topological number is n(t). Its eigenvectors 
\begin{equation}
\psi_\pm=\frac{1}{\sqrt{2}}\begin{pmatrix}
e^{i\phi(k)} \\ \pm1
\end{pmatrix},
\end{equation} 
where $\phi(k)\equiv\arctan{\frac{d_y(k)}{d_x(k)}}$. We assume that the parameters of Hamiltonian $t_1(t)=t_i$ and $n(t)=n_i$ at $t<0$, and after quench $t_1(t)=t_f$ and $n(t)=n_f$ at $t\geq0$. 
\begin{equation}
\abs{c_{+}(k)}^2=\abs{\bra{\psi_{-}(t_i,n_i,k)}\ket{\psi_+(t_f,n_f,k)}}^2=\sin^2{(\Delta\phi/2)},
\end{equation}
where $\Delta\phi=\phi_f-\phi_i$. Obviously, this result satisfies our conclusion. The numerical results for four different cases of this model are shown in Fig.\ref{ssh}.

